\documentstyle[amssymb,preprint,aps]{revtex}

\begin{document}
\author{A. Keren$^{*}$, P. Mendels$\dagger $ A. Kratzer$^{\ddagger }$, A.~Scuiller$%
^{\S }$, M.~Verdaguer$^{\S }$, Z. Slaman$^{*}$ and C.~Baines$^{\P }$}
\address{$^{*}$Technion-Israel Institute of Technology, Physics Department,\\
Haifa\\
32000, Israel\\
$^{\dagger }$Laboratoire de Physique des Solides, B\^{a}timent 510, URA2 CNRS%
\\
Universit\'{e}\\
Paris Sud, 91405 Orsay, France\\
$^{\ddagger }$Technische University Physics Department E15, James-Franck-Str,%
\\
D-85747 Garching, Germany\\
$^{\S }$Laboratoire de Chimie des M\'{e}taux de Transition, URA CNRS.\\
419, Universit\'{e}\\
Pierre et Marie Curie, 75252 Paris, France\\
$^{\P }$Paul Scherrer Institute, CH 5232 Villigen PSI, Switzerland}
\title{Magnetic quantum tunneling in the half-integer high spin molecule CrNi$_{6}$
: a $\mu $SR study}
\date{\today}
\maketitle

In high spin molecules metal ions are coupled by ferro or antiferromagnetic
short range interactions so that their magnetic moments are parallel or
antiparallel to each other at temperatures ($T$) much smaller than the
coupling constant $J$. This leads to a high spin ($S$) value in the ground
state. In the crystal lattice, the molecules are well separated from each
other and the active part of the Hamiltonian at $k_{B}T\ll J$ is ${\cal H}%
=-DS_{z}^{2}$, so that up and down spins ($S_{z}=\pm S$) states have
identical energies, and $S_{z}$ can escape from one state to the other via
either over-the-barrier motion, or tunneling. So far{\bf ,} the escape rate $%
\Gamma $ in HSM was found to have smooth $T$ dependence throughout the
cooling process from $DS^{2}\ll k_{B}T$ to $k_{B}T\ll DS^{2}$ \cite
{ThomasNature96,FriedmanPRL96}. In this case, the transition from the
over-the-barrier motion to tunneling, is referred to as second-order
transition \cite{ChudnovskyPRL97}. In the present work, we use the muon spin
relaxation technique to show evidence for temperature-independent tunneling
at $k_{B}T\ll DS^{2}$, in zero applied magnetic field (ZF), in the HSM CrNi$%
_{6}$. Moreover, the crossover from the classical to quantum regime is
sharp, and appears to be consistent with a first order transition of the
escape rate \cite{ChudnovskyPRL97}.

In a{\bf \ }first order transition, over-the-barrier motion is abruptly
replaced by tunneling between ground states, as the temperature is lowered
down to the barrier height.

The CrNi$_{6}$ molecule is described in details in the methods section. The
magnetization $M$ vs magnetic field $H$ measured at 6~K in this compound is
depicted in Fig.~\ref{Molecule}. $M$ displays a saturation value of $15$ $%
\mu _{\text{B}}$, and fits well to a Brillouin function of $S=15/2$. This $S$
value is expected from a Cr(III) ($S_{\text{Cr}}=3/2$) ion ferromagnetically
coupled to $6$ Ni(II) ($S_{\text{Ni}}=1$) ions. The ferromagnetic nature of
the interaction is easily rationalized in the frame of a localized electron
model, since the quasi-linear Cr-CN-Ni unit allows orthogonality between the
Cr(III) and Ni(II) wave functions.

Also shown in Fig.~\ref{Molecule} is the thermal variation of the molar
susceptibility as the product $\chi _{M}T$ vs $T$, which monotonously
increases from room temperature to 6K. The fit of the measured
susceptibility to the one expected from the Hamiltonian $H=-J\sum_{\text{Ni}}%
{\bf S}_{\text{Cr}}\cdot {\bf S}_{\text{Ni}}-D(S_{z}^{t})^{2}$, where $S^{t}$
is the total spin of the molecule, gives a good agreement when choosing a
mean $g$ value $=2$, a $J$ value $\approx 24$K, and $\left| D\right|
(S^{t})^{2}=4.2$ K \cite{MallahNATO} (see Fig.~\ref{Molecule}). The value of 
$\left| D\right| S^{2}$ is in agreement with ac-susceptibility measurements 
\cite{MallahJCS95}. $S^{t}=15/2$ is well separated in energy by 3J/2 (36K)
from an excited level where $S^{t}=13/2$\ , itself 3J/2 below a second level
where $S^{t}=11/2$, etc.

The $\mu $SR experiments presented here were performed in PSI after
preliminary investigation at ISIS using the longitudinal field (LF)
technique. In this technique one follows the time evolution of the spin
polarization $P_{z}(t)$ of a muon implanted in a sample, through the
asymmetry $A(t)\propto P_{z}(t)$ in the positron emission of the muon decay 
\cite{DalmasJPCM97}. In addition, an external field $H_{\text{L}}$ is
applied along the initial muon spin (longitudinal) direction which defines
the ${\bf z}$ axis. In view of our experimental results, it is worth
discussing the expected behavior of $P_{z}(t)$ in two cases: (I) the field
at the muon site is random and static or (II) the field is dynamically
fluctuating. In the static case, and when $H_{\text{L}}=0$, the muon
polarization relaxes, and the time scale for relaxation $[\tau ]$ is
determined by $(\gamma _{\mu }[B])^{-1}$ where $\gamma _{\mu }=85.162$%
{}~MHz/kG is the muon gyromagnetic ratio and $[B]$ is the scale of the local
magnetic field. However, not all muons will relax since there is a
statistical probability that some of them (effectively $1/3$) will reside in
a site where the field is parallel to the muon spin. Therefore, after a
minimum in $P_{z}(t)$, we expect $\lim_{t\rightarrow \infty }P_{z}(t)=1/3$ ,
a phenomenon called recovery \cite{DalmasJPCM97}. When the external field is
applied, and $[B]\ll H_{\text{L}}$, the vector sum of $H_{\text{L}}$ and the
internal field is nearly parallel to the muon spin, and we expect $%
\lim_{t\rightarrow \infty }P_{z}(t)=1$. This phenomenaon is called
decoupling \cite{DalmasJPCM97}. In the dynamical case the muon relaxation is
determined by the magnetic spectral density at the frequency $\omega _{\text{%
L}}=\gamma _{\mu }H_{\text{L}}$ corresponding to the muon Zeeman energy
levels (see below).

In Fig.~\ref{T50mK} we show the asymmetry, obtained in CrNi$_{6}$, in ZF, $%
T=50$~mK, and at various values of $H_{\text{L}}$. In the ZF case we find $%
\lim_{t\rightarrow \infty }A(t)=0$. This limit suggests that the field at
the muon site is not static. Moreover, the time scale for the muon
relaxation is $1$ $\mu $sec. If this relaxation was due to static fields it
would imply a local internal field in the order of tens of Gauss. Such a
small field would have decoupled [$A(t)=A(0)$] with $H_{\text{L}}=2$~kG or
higher, in contrast to the observed behavior. We thus conclude that the CrNi$%
_{6}$ molecules are dynamically fluctuating even at $T=50$~mK and in zero
applied magnetic field. We interpret these fluctuations as tunneling within
the ground spin state, between $S_{z}$ levels at the bottom of the
anisotropy well, since at 50~mK only these levels are populated. A similar
conclusion was reached by Vernier {\it et~al.} \cite{VernierPRB97} using
high frequency experiments.

In order to illuminate this conclusion we show in the inset of Fig.~\ref
{T50mK} a similar experiment in a powder of the Ising spin glass Fe$_{0.05}$%
TiS$_{2}$ ($T_{g}=15.5$~K) at $T=4$~K \cite{Fatih}. Here, as in the HSM, the
most important degree of freedom is $S_{z}$, and the classical ground state
is made of spins randomly oriented in either up or down directions. However,
in contrast to CrNi$_{6}$, we can clearly see in Fe$_{0.05}$TiS$_{2}$ a
recovery of the asymmetry. In addition, the time scale of muon relaxation in
Fe$_{0.05}$TiS$_{2}$ is an order of magnitude faster than in CrNi$_{6}$: yet
a field of 720~G is sufficient to nearly fully decouple the muon
polarization. Thus, Fe$_{0.05}$TiS$_{2}$ fulfills all our expectations from
a static magnet and emphasizes in contrast the dynamical nature of CrNi$_{6}$%
.

Next we determine the fluctuation rate $\nu $ (proportional to the escape
rate $\Gamma $), where $\nu $ is defined by $\left\langle
S^{t}(t)S^{t}(0)\right\rangle =\left\langle (S^{t})^{2}\right\rangle \exp
(-\nu t)$. Since at low temperatures the spins on the molecule are locked to
each other, the fluctuation rate of an individual spin should be identical
to that of the entire molecule. Therefore, the relaxation of the muon spin,
which is mostly coupled to the electronic spin closest to its stop site,
will be determined by $\nu $. A second important parameter is the root mean
square of the field at the muon site $\Delta $ $(=\gamma _{\mu }\left\langle
B_{i}^{2}\right\rangle ^{1/2})$ where $i$ is a spatial direction.{\bf \ }We
notice that the muon spin relaxation has a square-root exponential shape,
which could be explained by a distribution of $\Delta $, due to a
distribution of muon sites, according to $\rho (\Delta )=\sqrt{2/\pi }\frac{a%
}{\Delta ^{2}}\exp \left( -\frac{a^{2}}{2\Delta ^{2}}\right) $. Usually, one
can separate the two contributions to the muon relaxation rate by using the
LF data. Here the situation is more intricate as the impact of the applied
field on the tunneling rate is not evident and in a powder sample $H_{\text{L%
}}$\ also couples to $S_{x}$\ and $S_{y}$, and induces tunneling, or to $%
S_{z}$, and reduces tunneling (for most field values). In the absence of any
model, we assume that $\nu $\ is field independent for the fields ($g\mu
_{B}SH\lesssim DS^{2}$) used here, so as to get an estimate of $\nu $. These
assumptions lead to 
\begin{equation}
A(t)=A_{0}\exp \left[ -\sqrt{\frac{4a^{2}\nu t}{(\omega _{\text{L}}^{2}+\nu
^{2})}}\right] +A_{\text{b}}  \label{StrExp}
\end{equation}
where, $A_{0}$ is the initial asymmetry, and $A_{\text{b}}$ represent muons
that missed the sample \ \cite{KerenPRB94}. We fit our data to Eq.~\ref
{StrExp} using $A_{0}$, $A_{\text{b}}$, and $\nu $, as global parameters for
all data sets. We find reasonable agreement between theory and experiment
with a fluctuation rate $\nu =85(5)$ MHz and $a=10.0(5)$~MHz ($\sim 100$~G).
The fit, shown in Fig.~\ref{T50mK} by the solid lines, suggests that indeed
the external field might have only minimal impact on the powder averaged
tunneling rate at base temperature. One possible explanation for this is
that the interaction responsible for tunneling is much stronger than the
Zeeman coupling to the external field.

Now we discuss the temperature dependence of the muon relaxation. In Fig.~%
\ref{ZF} we depict the muon asymmetry in LF of 100~G and various
temperatures; the small field of 100~G is applied in order to decouple the
contribution from nuclear moments to the relaxation. We see that from $T=50$
to $6$~K the muon relaxation rate increases, but between $6$~K and $50$~mK
it remains constant (no change in the asymmetry). The fact that the
transition temperature is $\sim 6$~K reinforces our basic assumption that
the electronic spins determine the muon relaxation, since this temperature
is of the same order of magnitude as the energy barrier. The $T$ independent
fluctuation rate below $\sim 6$~K implies that these fluctuations stem from
QTM, as opposed to thermally activated, over-the-barrier motion.

We fit the data obtained at different temperatures to Eq.~\ref{StrExp}, with 
$\nu $ the only free parameter, and plot the result in Fig.~\ref{NuVsT}.
Clearly, the temperature dependence of $\nu $ changes abruptly at $T\sim 10$%
~K. Also shown in this figure by a solid line is a fit of $\nu $ to an
Arrhenius-like law 
\begin{equation}
\nu =\nu _{q}+\nu _{c}\exp (-U/T).  \label{escape}
\end{equation}
where $\nu _{q}=85$~$\mu $sec$^{-1}$. We find $\nu _{c}=12.0(3)$~nsec$^{-1}$%
, and $U=52(3)$~K. The large value of $U$ indicates that for $T>10$~K the
total spin states $S^{t}=13/2$ and $S^{t}=11/2$, become populated (recall
that $J_{\text{Ni-Cr}}=24$~K and that the gap between the lowest total spin
states is $36$~K). A comparison of Fig.~\ref{NuVsT} with ~Fig.~\ref{Molecule}
suggests that between $\sim 10$~K and $\sim 6$~K the molecules condense in a
total spin state $S^{t}=15/2$. The behavior below $\sim 6$~K suggests that
once the temperature becomes as low as the barrier height ($\left| D\right|
(S_{z}^{t})^{2}=4.2$ K), tunneling is taking place between the $S_{z}=\pm
15/2$ states; otherwise, it would have shown temperature dependence down to $%
\sim \left| D\right| \left[ (15/2)^{2}-(13/2)^{2}\right] =1.05$~K. This
behaviour appears closely related to a first order quantum-classical
transition of the escape rate \cite{ChudnovskyPRL97}. It should be pointed
out that the nature of the transition is inferred from a nearly zero field
experiment and is independent of the absolute value of $\nu $, which was
determined using the LF.

Our result renders CrNi$_{6}$ the first example of a high spin molecule
where the tunneling is temperature independent at low $T$, and the
transition from the classical activated behavior to the quantum one is sharp
and consistent with first order.

\newpage

{\bf Methods}

CrNi$_{6}$ -chemical formula [Cr\{(CN)Ni(tetren)\}$_{6}$](ClO$_{4}$)$_{9}$-
is obtained by reacting solutions of 0.1 molar potassium
hexacyanochromate(III) and 0.15 molar Ni(II) tetren perchlorate in a mixed
solvent water/acetonitrile. The solution is left to evaporate. After a few
weeks violet parallelepipedic single crystals appear which are collected,
filtered and dried in air. Single crystal X-ray diffraction shows that the
compound crystallises in the monoclinic system, P21/n space group, cell
parameters $a=15.375$~\AA , $b=24.280$~, $c=16.141$~\AA , $\alpha =90{%
{}^{\circ }}$, $\beta =115.580{{}^{\circ }}$, $\gamma =90{{}^{\circ }}$.
Disorder on the perchlorate anions and on the carbon framework of the
terminal amine ligands impedes a complete determination of the structure.
Nevertheless, it appears clearly that the system is made of heptanuclear
entities where the Cr(III) ion is surrounded by six cyanide ions, each
bonded to a Ni(II) tetren (see Fig.~\ref{Molecule}). The coordination sphere
of Cr(III) and Ni(II) can be described as a slightly distorted octahedral.
The Cr-CN-Ni distance is 5.23~\AA . The Cr-CN angle is $176.02{{}^{\circ }}$%
; the CN-Ni angle is $170.46{{}^{\circ }}$. The heptanuclear units are far
away from each other (nearest Cr-Cr distance = 14.77~\AA\ and nearest Ni-Ni
distance from neighbouring units = 8.84 \AA ).

\newpage

{\bf Acknowledgements }

We should like to thank T.~Mallah who first synthesized the compound and for
many discussions in the previous steps of the work. We should also like to
thank the technical staff of both ISIS and PSI and especially P.~J.~King and
A.~Amato for their continuous assistance. We are grateful to
E.~M~Chudnovsky, G. Bellessa, and P.~Stamp for many suggestions and
discussions. These experiments were supported by the European Union through
its Training and Mobility of Researchers Program for Large Scale Facilities,
and by the Israeli Academy of Science.

\begin{figure}[tbp]
\caption{Thermal dependence of the molar susceptibility $\chi_M T$ vs $T$,
and Magnetisation vs H at 6K (symboles). The fits (lines) are described in
the text. Also shown is the CrNi$_6$ molecule. }
\label{Molecule}
\end{figure}

\begin{figure}[tbp]
\caption{Muon asymmetry verseus time in CrNi$_{6}$ at $T=50$~mK and various
applied longitudinal magnetic fields. This figure demonstrates the dynamical
nature of the electronic spins (see text). For comparison the inset shows
the muon asymmetry in the Ising spin glass Fe$_{0.05}$TiS$_{2}$ ($T_{g}=15.5$%
~K) at $T=4$~K where these spins are static.}
\label{T50mK}
\end{figure}

\begin{figure}[tbp]
\caption{Muon asymmetry versus time in CrNi$_{6}$ at $H_{\text{L}}=100$~G
and various temperatures (symbols). The solid line is a fit to Eq.~1.}
\label{ZF}
\end{figure}

\begin{figure}[tbp]
\caption{Temperature dependence of the molecular spin fluctuation rate. The
solid line is a fit to Eq.~2. }
\label{NuVsT}
\end{figure}

\end{document}